\documentclass[twocolumn]{aastex7}

\usepackage{amsmath}

\begin{document}

\title{Multi-Wavelength Identification of a Luminous Mid-Infrared Supernova Powered by Circumstellar Interaction with Binary-Driven Pre-supernova Mass Loss}


\author[orcid=0000-0002-6765-8988]{Kohki Uno}
\affiliation{Department of Astronomy and Columbia Astrophysics Laboratory, Columbia University, New York, NY, USA}
\email[show]{ku2204@columbia.edu}  

\author[orcid=0000-0002-8989-0542]{Kishalay De} 
\affiliation{Department of Astronomy and Columbia Astrophysics Laboratory, Columbia University, New York, NY, USA}
\affiliation{Center for Computational Astrophysics, Flatiron Institute, New York, NY, USA}
\email{kd3038@columbia.edu}

\author[orcid=0000-0001-6331-112X]{Geoffrey Mo} 
\affiliation{Division of Physics, Mathematics and Astronomy, California Institute of Technology, Pasadena, CA 91125, USA}
\affiliation{The Observatories of the Carnegie Institution for Science, 813 Santa Barbara St, Pasadena, CA 91101, USA}
\email{gmo@caltech.edu}

\author[0000-0003-2758-159X]{Viraj Karambelkar}
\email{vk2588@columbia.edu}
\altaffiliation{NASA Hubble Fellow}
\affiliation{Columbia University, 538 West 120th Street 704, MC 5255, New York, NY 10027}
\email{vk2588@columbia.edu}

\author[orcid=0000-0002-6823-2073]{Kaitlyn Shin} 
\affiliation{Division of Physics, Mathematics and Astronomy, California Institute of Technology, Pasadena, CA 91125, USA}
\email{kaitshin@caltech.edu}

\author[orcid=0000-0002-4223-103X]{Christoffer Fremling} 
\affiliation{Caltech Optical Observatories, California Institute of Technology, Pasadena, CA 91125, USA}
\affiliation{Division of Physics, Mathematics and Astronomy, California Institute of Technology, Pasadena, CA 91125, USA}
\email{fremling@caltech.edu}

\author[orcid=0000-0003-1546-6615]{Jesper Sollerman} 
\affiliation{The Oskar Klein Centre, Department of Astronomy, Stockholm University, AlbaNova, SE-10691 Stockholm, Sweden}
\email{jesper@astro.su.se}

\author[orcid=0000-0002-1125-9187]{Daichi Hiramatsu}
\affiliation{Department of Astronomy, University of Florida, Bryant Space Science Center, Gainesville, FL 32611-2055, USA}
\email{dhiramatsu@ufl.edu}  

\author[orcid=0000-0001-5754-4007]{Jacob E. Jencson}
\affiliation{Caltech/IPAC, Mailcode 100-22, Pasadena, CA 91125, USA}
\email{jjencson@ipac.caltech.edu}


\begin{abstract}
Terminal mass-loss in massive stars is commonly reconstructed from observations of circumstellar interaction in core-collapse supernovae (CCSNe); however, such signatures can be easily missed depending on the phase when they appear. In this paper, we present a multi-wavelength study of SN\,2022yyz, a nearby CCSN selected by its luminous late-time mid-infrared (MIR) excess in {\it NEOWISE} at $> 500$\,days after the discovery. Classified as a Type II SN at discovery, it reached a peak bolometric luminosity of $1.3^{+0.7}_{-0.4}\times10^{43}$ erg s$^{-1}$ before fading away with a light-curve drop at $\sim 300$\,days. We present late-time {\it HST}/WFC3 ultraviolet (UV) imaging and Keck/LRIS spectroscopy at $\sim 1200$\,days, which reveal clear signs of circumstellar-medium (CSM) interaction via a UV excess consistent with a strong \ion{Mg}{2} contribution and a flat-topped H$\alpha$ profile. We show that the bolometric light curve and the late-time UV luminosity can be explained using a semi-analytic CSM interaction model involving a broken power-law density profile: with $\sim 2.7~M_\odot$ of nearby, dense CSM within a radius of $\sim 10^{16}$ cm surrounded by wind-like outer CSM with a mass-loss rate of $\sim 10^{-4}\,{\rm M_\odot\,yr}^{-1}$. The broken density profile indicates $\gtrsim 100\times$ mass-loss rate enhancement starting a few hundred years before explosion, pointing to a possible common-envelope ejection in a binary system. Despite being overlooked for follow-up at such proximity, our results demonstrate that IR photometric selection methods provide a powerful way of mapping terminal mass-loss in massive stars -- laying the groundwork for the {\it Roman} Space Telescope surveys.
\end{abstract}




\section{Introduction} \label{sec:intro}

Massive stars ($\gtrsim 8M_\odot$) are known to undergo intense mass loss prior to their terminal explosions as core-collapse supernovae \citep[CCSNe;][for review]{Smith_2014_ARAA_52_487S}. Such pre-SN mass loss provides a direct probe of the poorly understood final stages of massive stellar evolution, when processes such as unstable late-stage nuclear burning, wave-driven energy transport \citep{Quataert_2012_MNRAS_423L_92Q, Fuller_2017_MNRAS_470_1642F, Wu_2021_ApJ_906_3W}, and binary interaction \citep{Sana_2012_Science_337_444S, Pejcha_2016_MNRAS_455_4351P} can substantially alter the progenitor structures and their environments. Observational evidence for such pre-SN mass loss was historically first inferred through signatures of interaction between SN ejecta and circumstellar material (CSM), commonly manifesting as narrow ($\lesssim 1000$ km s$^{-1}$) emission lines in CCSN spectra \citep{Smith_2017_snhandbook_403S}.

The wide dynamic range of timescales for pre-SN mass loss implies that such spectroscopic signatures manifest on a range of temporal scales. Mass loss occurring in the very final days to weeks before explosion gives rise to confined CSM close to the progenitor, whose interaction with the SN ejecta produces strong but fleeting highly ionized emission features, a technique referred to as ``flash spectroscopy'' \citep{GalYam_2014_Nature_509_471G, Yaron_2017_NatPh_13_510Y, Bruch_2021_ApJ_912_46B}. On the other hand, sustained mass loss starting years to decades before explosion gives rise to the class of Type IIn SNe \citep{Smith_2017_snhandbook_403S}, which exhibit narrow hydrogen emission lines and long-lasting luminosity powered by sustained CSM interaction \citep[][for review]{Filippenko_1997_ARAA_35_309F}. In either case, evidence for such dense nearby CSM is commonly limited to the SN appearance before and around peak light, where the SN brightness makes it amenable to routine spectroscopy.

In cases where the pre-SN mass loss is terminated shortly prior to explosion, evidence for nearby CSM may be easily missed by early-time observations. Episodic mass loss occurring $\sim 10^2$--$10^3$ yr before explosion is expected to produce CSM enhancements at $\sim 10^{16}$--$10^{17}$\,cm; interaction with such distant, detached CSM would occur only at late phases ($\gtrsim 1$--$2$ yr after explosion), when the resulting emission is expected to be faint and difficult to monitor. While such searches have been historically limited by observational capabilities, evidence for their existence is now clearly emerging from individual detailed studies in X-ray, radio, and optical/UV bands \citep[e.g.,][]{Milisavljevic_2015_ApJ_815_120M, Margutti_2017_ApJ_835_140M, Graham_2019_ApJ_871_62G, Fremling_2026arXiv260329043F, Baer-way_2026arXiv260705500B}.  More importantly, episodic mass-loss on these timescales poses an important challenge to models of massive-star evolution, with most stellar evolution models falling short of explaining eruptive mass loss on such long timescales \citep{Smith_2014_ARAA_52_487S, Woosley_2017_ApJ_836_244W} that far surpass those expected for terminal nuclear processes in stellar interiors \citep[$\lesssim 10$\,yr; e.g.,][]{Quataert_2012_MNRAS_423L_92Q,Shiode_2014_ApJ_780_96S, Fuller_2017_MNRAS_470_1642F, Wu_2021_ApJ_906_3W}.

Earlier studies with {\it Spitzer Space Telescope} demonstrated the utility of mid-infrared (MIR) observations in searching for late-time CSM interaction -- particularly in Type IIn SNe \citep{Fox_2011_ApJ_741_7F, Fox_2013_AJ_146_2F}. Because CSM interaction ubiquitously emits in soft X-ray and UV bands \citep{Chevalier_2017_snhandbook_875C, Fransson_2014_ApJ_797_118F, Dessart_2022_AA_660_L9D} where dust is effectively opaque, excess MIR emission at late phases (exceeding radioactive decay) becomes an unmistakable signature of extended CSM shells \citep[e.g.,][]{Tinyanont_2016_ApJ_833_231T, Szalai_2019_ApJS_241_38S, Szalai_2021_ApJ_919_17S}. Although such {\it Spitzer} studies were limited to a sample of nearby galaxies, recent systematic searches with the Wide-field Infrared Survey Explorer ({\it WISE})/{\it NEOWISE} all-sky survey \citep{Wright_2010_AJ_140_1868W, Mainzer_2014_ApJ_792_30M} have enabled the first systematic searches for SN interaction with distant/detached CSM shells \citep{Myers_2024_ApJ_976_230M, Mo_2025_ApJL_980_33M}. The combination of {\it NEOWISE} data with recent controlled samples of nearby CCSNe now indicate that $\sim 3-5\%$ of CCSNe without clear interaction signatures at early phases exhibit MIR rebrightening at $\gtrsim 1000$ days after explosion, implying that distant/detached CSM shells are at least as common as the more extensively studied class of Type IIn SNe \citep{Myers_2024_ApJ_976_230M}.

Despite its diagnostic potential, MIR photometry alone does not uniquely determine the physical properties of the CSM, such as its mass, geometry and covering fraction, which are critical to establishing the nature of the pre-SN mass loss. Multi-wavelength follow-up is necessary to establish whether MIR-selected events are indeed powered by ongoing CSM interaction and to quantify their mass and energy budgets. In particular, UV observations provide a sensitive probe of the kinetic energy dissipation that emerges as electromagnetic radiation, as theoretical models predict that a substantial fraction of the radiation from late-time CSM interaction appears in strong UV emission of \ion{Mg}{2} $\lambda2800$ \citep{Dessart_2022_AA_660_L9D}. Furthermore, recent {\it HST} observations have demonstrated the utility of the UV emission as a probe of late-time CSM interaction \citep{Fremling_2026arXiv260329043F, Bostroem_2026_ApJ_1004_23B}.
On the other hand, optical spectroscopy offers a complementary diagnostic of the kinematics and geometry of the CSM, with line profiles constraining the velocity field, geometry, and density structure of the interaction region.

To this end, we are carrying out a large systematic survey combining the complementary diagnostics provided by {\it NEOWISE} MIR observations, UV observations from the {\it Hubble Space Telescope (HST)} and ground-based optical spectroscopy with the Keck-I telescope. In this paper, we present a multi-wavelength analysis of SN 2022yyz \citep{Forster_2022_TNSTR_3149_1F}, a nearby CCSN that was classified as a Type II supernova \citep{Zhai_2022_TNSCR_3243_1Z} with early spectroscopy; however, we subsequently identified this source for additional follow-up based on its luminous late-time MIR excess in {\it NEOWISE} data. We present the first results from this multi-wavelength survey via detailed characterization of this event. This paper is structured as follows. Section~\ref{sec:observations} describes the observations and data analysis. Section~\ref{sec:obsprop} presents the observational properties of SN\,2022yyz. Section~\ref{sec:modeling} models the spectral energy distributions, the late-time H$\alpha$ profile, and the bolometric light curve, from which we derive the CSM structure and the mass-loss history. We discuss the implications and summarize our conclusions in Section~\ref{sec:discussion}.


\section{Observations} \label{sec:observations}

\begin{figure*}[htb]
\centering
\epsscale{1.17}
\gridline{
  \fig{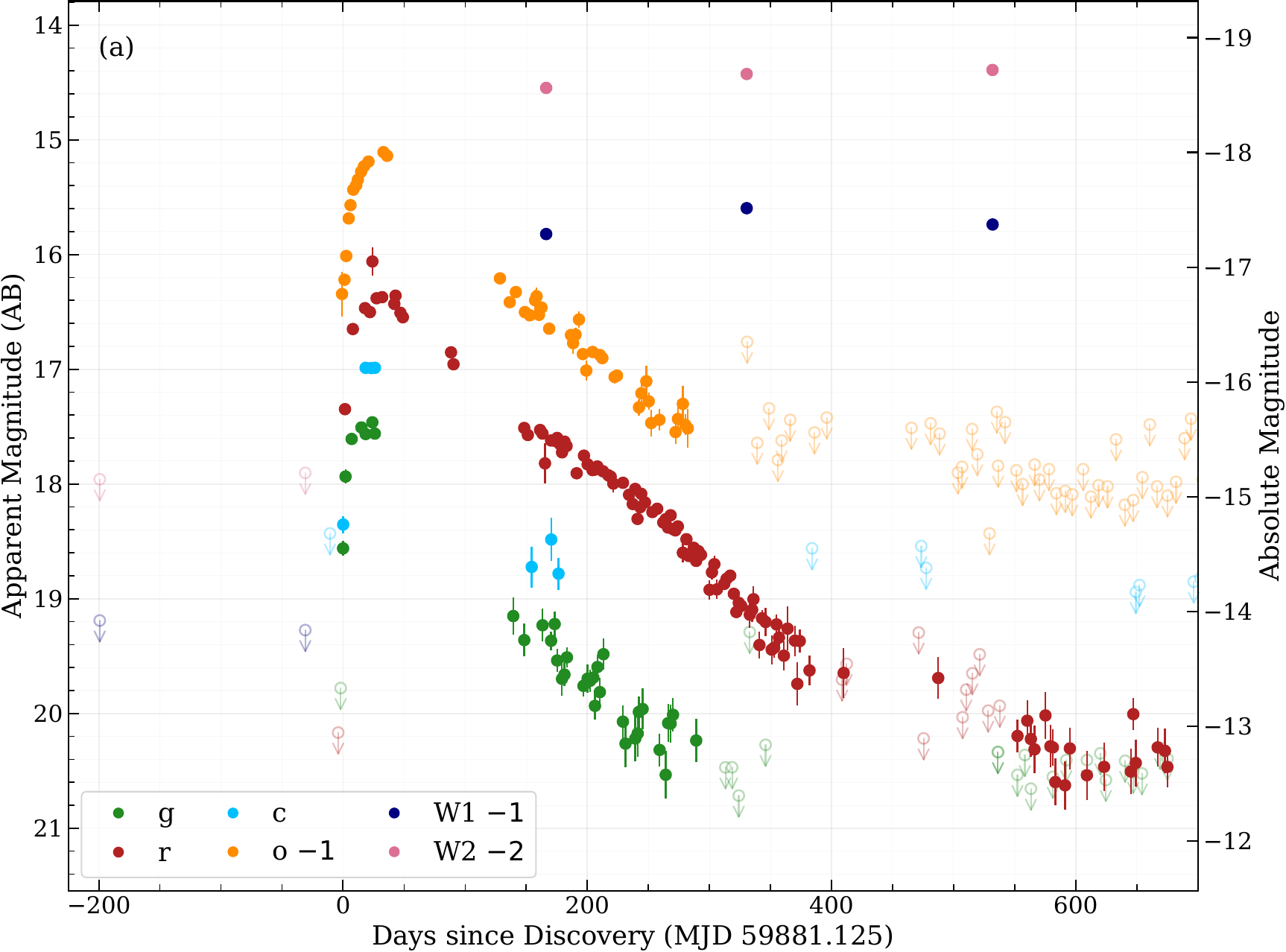}{0.56\textwidth}{}
  \fig{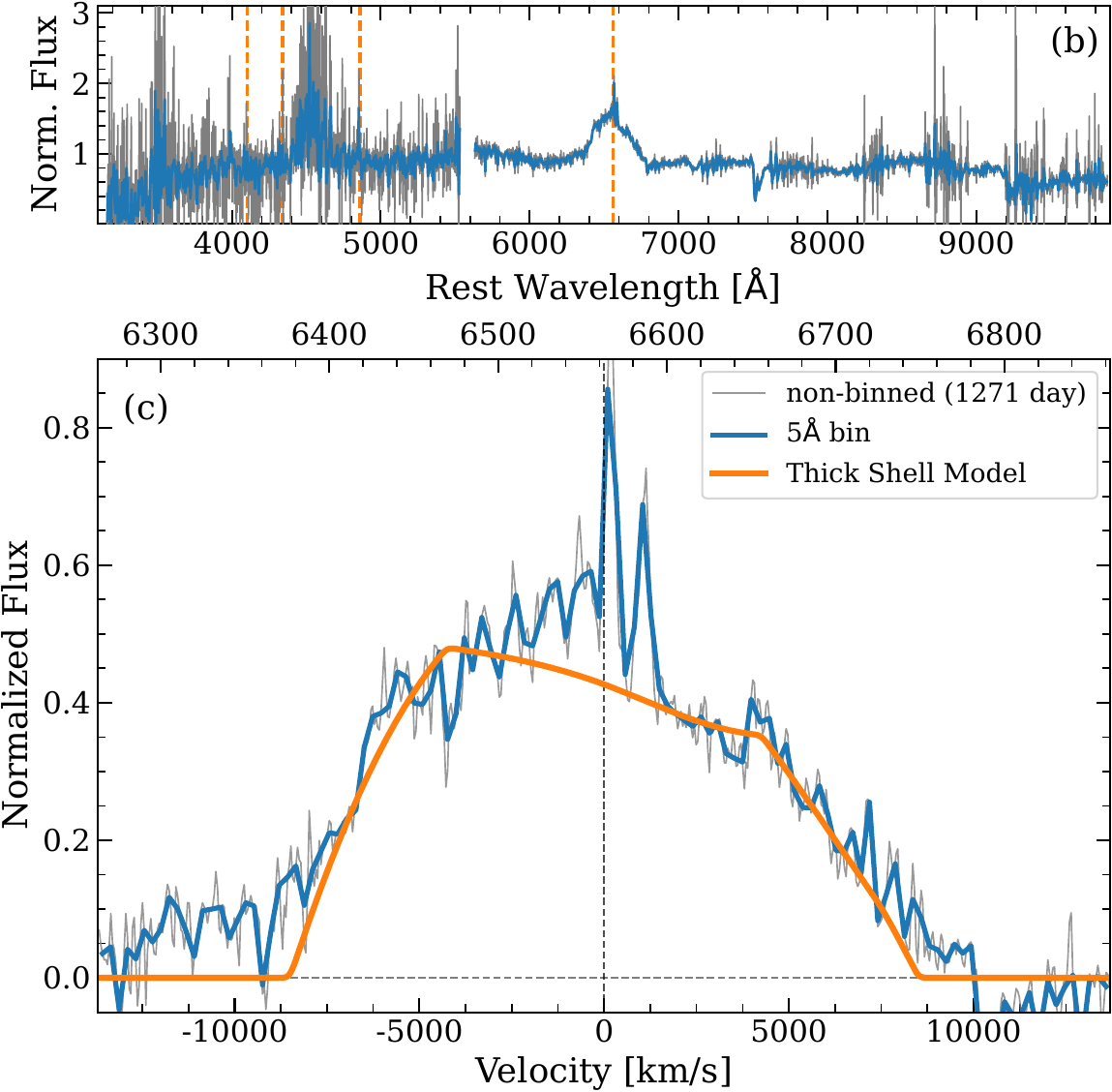}{0.42\textwidth}{}
}
\caption{
Panel (a): Multi-band light curve of SN 2022yyz. Filled circles show detections with $1\sigma$ errors in the ZTF $g$ (green) and $r$ (red) bands, the ATLAS $c$ (cyan) and $o$ (orange) bands, and the WISE $W1$ (navy) and $W2$ (pink) bands, with offsets by the values shown in the legend. Open circles with downward arrows denote $5\sigma$ upper limits (non-detections). 
The left axis gives the apparent AB magnitude corrected for the Milky Way extinction, and the right axis the corresponding absolute magnitude. Note that the light curves are not corrected for the host-galaxy extinction.
Panel (b): Normalized spectrum of SN 2022yyz obtained with Keck/LRIS at 1271 days after the discovery. The gray spectrum shows the unbinned spectrum, while the blue spectrum shows the spectrum binned to 5\,{\AA}. The orange dashed lines indicate the Balmer series. The spectral gap around 5500\,{\AA} corresponds to the transition between the red and blue arms of the D560 dichroic. The feature around 4500\,{\AA} is a noise artifact.
Panel (c): Normalized H$\alpha$ profile plotted in velocity space. The orange line shows the thick-shell model profile described in Sec.~\ref{subsec:modeling_spec} and Appendix~\ref{app:line_profile}. Note that the narrow feature at $\sim 1000$\,km\,s$^{-1}$ corresponds to [\ion{N}{2}]\,$\lambda6584$. Both narrow features likely originate from the host background (Section~\ref{subsec:obsprop_spec}), while the broad H$\alpha$ emission arises from the SN.
}
\label{fig:optical}
\end{figure*}

\subsection{Optical Photometry} \label{subsec:optical_photometry}

SN 2022yyz was discovered by ZTF on 2022 October 29 \citep[MJD 59881.125,][]{Forster_2022_TNSTR_3149_1F} at (RA, Dec) $=$ (19:07:01.553, +28:59:50.09) in UGC 11404 at $z=0.0122$\footnote{independent discovery was reported by XOSS (TNS Astronomical Transient Report No. 163061 by Mi Zhang)}, and a redshift-independent distance for the host galaxy UGC~11404 is reported as $D_{\mathrm{L}} \approx 41.875 \pm 1.596$ Mpc \citep[e.g.,][]{Tully_2016_AJ_152_50T} \footnote{the luminosity distance corresponding to the redshift is evaluated as $50.4 \pm 1.5$ Mpc, adopting $H_{0}=73.2 \pm 2.3$ km s$^{-1}$ Mpc$^{-1}$ \citep{Burns_2018_ApJ_869_56B}, $\Omega_{\mathrm{M}}=0.27$, and $\Omega_{\mathrm{\Lambda}}=0.73$.}. In this paper, we adopt $D_{\mathrm{L}} = 41.875$ Mpc. It was classified as a Type II SN \citep{Zhai_2022_TNSCR_3243_1Z}, and was also included in the ZTF Type II SN sample \citep{Das_2025PASP..137d4203D}. The classification spectrum showed an unresolved hydrogen emission line with a velocity width of at most a few $\times 100$ km s$^{-1}$, and a strongly reddened continuum. We obtain optical light curves from public archives using the ZTF forced-photometry service \citep{Masci_2019_PASP_131a8003M} and the ATLAS forced-photometry server \citep{Tonry_2018_PASP_ATLAS}. We correct the photometry for Milky Way reddening of $A_{V} = 0.643$ mag \citep{Schlafly_2011_ApJ_737_103S} by adopting the \citet{Fitzpatrick_2007_ApJ_663_320F} extinction law with $R_{\mathrm{V}} = 3.1$. 

Even after correcting for Milky Way extinction, SN 2022yyz exhibits a significantly reddened light curve, with an observed peak color of $g-r \approx 0.9$ mag (see Figure~\ref{fig:optical}), suggesting additional dust extinction from the host galaxy. To estimate the host-galaxy extinction, we assume several plausible intrinsic peak colors for (interacting) SNe II: $(g-r)_{\mathrm{int}} = -0.1$, $0.0$, and $+0.1$ mag \citep[see; e.g.,][]{Galbany_2016_AJ_151_33G, deJaeger_2018_MNRAS_476_4592D, Nyholm_2020_AA_637_A73N}. Given the observed peak color, these intrinsic colors correspond to the color excesses of $E(g-r) \approx 1.0$, $0.9$, and $0.8$ mag, respectively. Assuming the same extinction law as the galactic extinction with $R_V = 3.1$ \citep{Fitzpatrick_2007_ApJ_663_320F}, these color excesses imply host-galaxy extinctions of $A_{V,\mathrm{host}} \approx 2.8$, $2.5$, and $2.2$ mag, respectively.

\subsection{Infrared Photometry} \label{subsec:IR_photometry}

We performed point-spread-function photometry at the source position in the {\it NEOWISE} difference images in the $W1$ (3.4 $\mu$m) and $W2$ (4.6 $\mu$m) bands, obtained with {\it WISE}. We followed the photometric method described by \citet{De_2020_PASP_132b5001D, Myers_2024_ApJ_976_230M}. Although classified as a Type II SN, the source exhibits a luminous, brightening MIR counterpart starting $\gtrsim 5$ months after discovery, indicating an overlooked event with strong CSM interaction (Figure~\ref{fig:optical}a). For subsequent modeling of the MIR photometry, extinction due to dust in both the Milky Way and the host galaxy is assumed to be negligible. 

Because the {\it NEOWISE} photometry only extends until the end of the mission in July 2024, we obtained additional ground-based near-IR observations on the Magellan Baade telescope. On UT 2025 May 13 (MJD 60808.3, 927 days after discovery), we obtained near-IR (NIR) images with the FourStar camera \citep{Persson_2013_PASP_125_654P}. The observations were obtained as dithered sequences of exposures amounting to a total exposure time of 917\,s, 524\,s and 437\,s in $J$, $H$ and $K_s$ bands, respectively. The data were reduced with flat-fielding, sky subtraction and aperture photometry calibrated to the 2MASS catalog as described in \citet{De_2020_PASP_132b5001D}. The SN is clearly detected in all three bands as $m_{J} = 20.41 \pm 0.03$, $m_{H} = 19.96 \pm 0.12$, and $m_{K_s} = 18.61 \pm 0.11$\,AB\,mag without the host-extinction correction. A second epoch of FourStar imaging was obtained on UT 2026 June 2 (MJD 61193.4, 1312 days), yielding $m_H = 20.27 \pm 0.11$ and $m_{K_{\rm s}} = 19.35 \pm 0.06$\,AB\,mag, without correction for host-galaxy extinction. The total exposure times are 786\,s and 524\,s in $H$ and $K_s$ bands, respectively. No $J$-band observation was obtained at this epoch (see also Appendix \ref{app:photo_tables}).

\subsection{Ultraviolet Photometry} \label{subsec:UV_photometry}

\begin{figure*}[htb]
\centering
\plotone{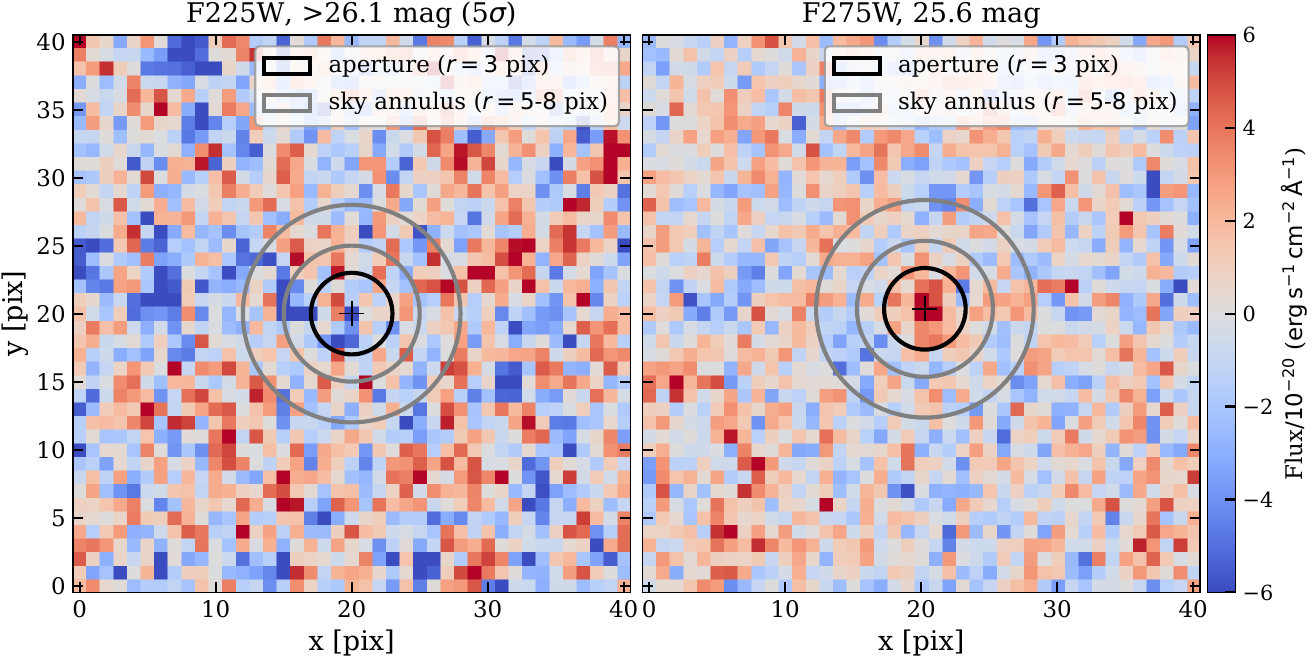}
\caption{
Cutout images in the F225W (left) and F275W (right) bands. The black and gray regions indicate the source aperture and sky annulus used for photometry, respectively. A point source is visible in the F275W image.
}
\label{fig:uv}
\end{figure*}

SN\,2022yyz was included in our {\it HST} follow-up program for CCSN candidates with distant/detached CSM shells identified from {\it NEOWISE} data. On 2026 March 27 (MJD 61126.0, 1245 days), SN\,2022yyz was observed with {\it HST}/WFC3 UVIS imaging in the F275W and F225W filters under program GO-18016 (PI: De\footnote{\url{https://www.stsci.edu/hst-program-info/program/?program=18016}}). For each filter, the observations consisted of two dithered 550\,s exposures. We performed aperture photometry with the aperture size of $3$\,pix (see Figure~\ref{fig:uv}) on the drizzled DRC images, which are geometrically corrected, drizzled images produced by combining the individual dithered exposures. Figure~\ref{fig:uv} shows the detection of a point source with $m_{\mathrm{F275W}} = 25.6 \pm 0.15$\,mag in F275W, while only an upper limit of $m_{\mathrm{F225W}} > 26.1$\,mag ($5\sigma$) was obtained in F225W. To verify the F275W detection, we examined the two individual FLC exposures, which are calibrated but retain the native geometric distortion. We detected a point source at the same position in both frames, with signal-to-noise ratios of $S/N \approx 2.9$ and $4.5$, respectively\footnote{We examined \textit{Swift}/UVOT archival data for SN~2022yyz obtained in April and May 2025. The source was not detected, providing 5$\sigma$ upper limits for UV detections: $m_{\mathrm{UVW2}} > 21.8$, $m_{\mathrm{UVM2}} > 21.1$, and $m_{\mathrm{UVW1}} > 20.8$\,AB\,mag, which are consistent with our \textit{HST} detection.}  (see also Appendix \ref{app:photo_tables}).

\subsection{Optical Spectroscopy} \label{subsec:optical_spectroscopy}

There is no publicly reported optical spectroscopy of SN\,2022yyz following its initial classification spectrum as a Type II supernova. On 2026 April 22 (MJD 61152.5, 1271 days after discovery), we obtained an optical spectrum of SN\,2022yyz with Keck/LRIS \citep{Oke_1995_PASP_107_375O} as part of a program to constrain the CSM geometry around {\it NEOWISE}-identified interacting SNe (Program 2026A/N194; PI: Uno). We used the D560 dichroic mirror, the 400/8500 grating, and the 400/3400 grism, covering $3000$--$10000$~\AA\ with a spectral resolution of $R=\lambda/\Delta \lambda \approx 1000$. The exposure times were $550$\,s $\times$ 6 for the red arm and $1200$\,s $\times$ 3 for the blue arm. For wavelength calibration, we used Hg/Ne/Ar/Cd/Zn arc lamps. The data were reduced with PypeIt\footnote{\url{https://pypeit.readthedocs.io/en/stable/index.html}} \citep{pypeit:joss_pub, pypeit:zenodo}, a semi-automated Python-based package for reducing astronomical spectroscopic data. The reduced spectrum is shown in Figure~\ref{fig:optical}b. We clearly detect prominent, broad $\mathrm{H\alpha}$ emission, while no other Balmer lines or continuum emission are detected, tentatively suggesting that SN\,2022yyz remains driven by ongoing circumstellar interaction $\sim 3.5$\,yr after discovery.


\section{Observational Properties} \label{sec:obsprop}

\subsection{Light Curves} \label{subsec:obsprop_optical}

Figure~\ref{fig:optical}a shows the optical and MIR light curves before the host-galaxy extinction correction. The optical light curves rise to peak within $\sim 30$\,days, and then monotonically decline without a clear plateau phase. The decline rates of SN\,2022yyz are $\sim 1.0$\,mag/50\,days for the $g$-band, and $\sim 0.6$\,mag/50\,days for the $r$-band, which are similar to those of Type IIL-like SNe \citep[e.g.,][]{Faran_2014_MNRAS_445_554F}. After correcting for the host-galaxy extinction, SN 2022yyz reaches a peak absolute magnitude of $M_{r} \approx -18.7 \pm 0.2$\,mag, placing it at the luminous end of the SNe IIL \citep[e.g.,][]{Faran_2014_MNRAS_445_554F, Gall_2015_AA_582_A3G} and within the typical luminosity range of SNe IIn \citep{Nyholm_2020_AA_637_A73N, Hiramatsu_2026_ApJ_1005_82H}.

Integrating blackbody fits to the extinction-corrected $g$- and $r$-band photometry, SN\,2022yyz reached a peak bolometric luminosity of $1.3^{+0.7}_{-0.4}\times10^{43}$\,erg\,s$^{-1}$ before fading away with a light-curve drop at $\sim 300$\,days (see also Section~\ref{subsec:modeling_Lbol} and Figure~\ref{fig:lc_model}). On the other hand, the MIR light curves place deep pre-explosion limits of $m_{\mathrm{W1}} > 19.28$\,mag and $m_{\mathrm{W2}} > 18.91$\,mag on MJD 59850 (31 days before the discovery), and then they show a luminous and nearly constant MIR excess, with $m_{\mathrm{W1}} \approx 16.6$--$16.8$\,mag and $m_{\mathrm{W2}} \approx 16.4$--$16.5$\,mag at the three {\it NEOWISE} epochs (166, 331, and 532 days). The color evolution in the MIR band provides insights into the powering source: The $W1-W2$ color becomes bluer by $0.1$ mag between epochs 1 and 2, favoring continuous energy injection from CSM interaction rather than a monotonically fading heating source.

\subsection{Ultraviolet Detection} \label{subsec:obsprop_uv}

After correcting for the Milky Way and host-galaxy extinction ($A_{\mathrm{F275W}} \approx 6.1$--$7.3$\,mag across the three host-extinction cases), the F275W detection of $m_{\mathrm{F275W}} = 25.6$\,mag corresponds to an absolute magnitude of $M_{\mathrm{F275W}} \approx -14.8$ to $-13.6$\,mag, or a band luminosity of $L_{\mathrm{F275W}} = W_{\mathrm{eff}}L_{\lambda} \approx 3.6^{+2.6}_{-1.5} \times 10^{40}$ erg s$^{-1}$, where $W_{\mathrm{eff}}$ is the band width. No source is detected in F225W down to $m_{\mathrm{F225W}} > 26.1$\,mag. These two bands are sensitive to different signatures of CSM interaction: F275W traces the \ion{Mg}{2} $\lambda 2800$ emission predicted by numerical models of late-time CSM interaction \citep{Dessart_2022_AA_660_L9D}, while F225W probes the neighboring UV continuum. The detection in F275W together with the non-detection in the adjacent F225W continuum band indicates that the UV emission is dominated by line emission rather than continuum, providing independent evidence for ongoing CSM interaction at 1245 days while disfavoring emission from an underlying star-forming region.

Late-time interaction models \citep{Dessart_2022_AA_660_L9D} predict that the dissipated kinetic energy is reprocessed into UV line emission with a conversion efficiency of $60$--$100\%$ at these phases. Adopting this range, the F275W luminosity implies a bolometric luminosity of $L_{\mathrm{bol}} \approx 4.5^{+6.0}_{-2.5} \times 10^{40}$\,erg\,s$^{-1}$. Because the optical emission has faded below detectability by this epoch, this UV-based estimate provides the only constraint on the bolometric output at $\sim 1245$\,days, corroborating the drop in the bolometric light curve at $\approx 300$\,days and anchoring the late-time evolution used in our CSM interaction modeling (Section~\ref{subsec:modeling_Lbol}; Figure~\ref{fig:lc_model}).

\subsection{Spectral Profile} \label{subsec:obsprop_spec}

We show an enlarged view of the detected $\mathrm{H\alpha}$ profile in velocity space (Figure~\ref{fig:optical}c). The emission line exhibits a broad component extending to $\sim 8500$\,km\,s$^{-1}$ without a P-Cygni profile, together with a narrow component centered at the rest wavelength. The broad component shows a flat-topped profile with suppression on the red side. A flat-topped profile is a natural signature of emission from a detached, expanding shell \citep{Jerkstrand_2017_snhandbook_795J}, while the red-side suppression indicates absorption within or interior to the emitting region \citep[e.g.,][]{Smith_2008_ApJ_686_467S}. The broad component is naturally interpreted as emission from the shocked, swept-up shell formed by the ejecta--CSM interaction. On the other hand, the narrow component is not fully resolved and is accompanied by [\ion{N}{2}]\,$\lambda6584$ of comparable width; both narrow features are spatially extended along the slit, and thus they likely originate from the background.

\section{Modeling} \label{sec:modeling}

\subsection{Spectral Energy Distributions} \label{subsec:modeling_sed}

\begin{figure*}[htb]
\centering
\epsscale{1.17}
\plotone{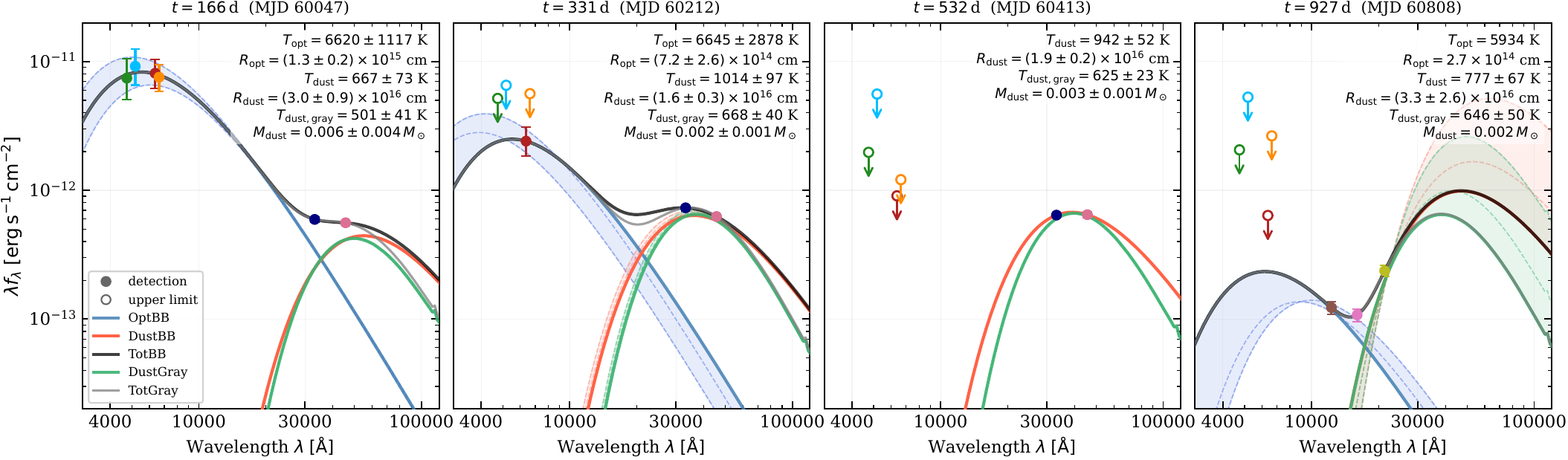}
\caption{
SEDs at $t=166$, $331$, $532$, and $927$\,days after discovery. The filled circles are detections (ZTF $g$/$r$, ATLAS $c$/$o$, {\it WISE} $W1$/$W2$, and FourStar $J$/$H$/$K_{\rm s}$), and the open circles with arrows are the upper limits. All photometric measurements are corrected for the Milky Way and host-galaxy extinction, except for the {\it WISE} $W1$/$W2$ data, for which no extinction correction is applied. The error bars include the photometric uncertainty and the spread among the host-extinction scenarios. The blue, red, and black lines show the optical blackbody, dust blackbody, and total two-component blackbody SEDs, while the green and gray lines show a graybody model for graphite grains with $a=0.1\,\mu$m fit to the IR excess and the corresponding total SED. The shaded regions show the host-extinction uncertainty. In the third epoch, only the dust components are shown, since only optical upper limits are available. The best-fit parameters for the fiducial host-extinction case: $(g-r)=0.0$\,mag, are shown in each panel. In the fourth epoch, the optical bands provide only upper limits and the shortest-wavelength detection is the $J$ band, so the optical blackbody component is essentially unconstrained at this epoch. The corresponding parameters are shown for illustration only and no uncertainties are quoted.
}
\label{fig:sed}
\end{figure*}

Figure~\ref{fig:sed} shows the spectral energy distributions (SEDs) for the all-extinction-corrected data at the four epochs for which sufficient photometry is available to constrain the SED, including the NIR detections at 927 days. We do not use the second FourStar epoch at 1312 days because only two NIR bands were obtained.
We fit the SEDs with two blackbody components, representing the optical (interaction-powered) emission and dust emission. Using a blackbody with a temperature $T_{\mathrm{BB}}$ and radius $R_{\mathrm{BB}}$, the observed flux is estimated as follows:
\begin{equation}
    f_{\lambda} = \pi B_{\lambda}(T_{\mathrm{BB}}) \frac{R_{\mathrm{BB}}^{2}}{D_{\mathrm{L}}^{2}},
    \label{eq:bb}
\end{equation}
where $B_{\lambda}$ is the Planck function. Note that we neglect time delay between the optical and MIR emission components, since the light crossing time to the dust shell ($\sim 10^{16}$\,cm) is a few days, which is sufficiently shorter than the luminosity-evolution timescale. The SED evolution indicates that the optical component fades with time, while the dust component comes to dominate the SED. The dust component is characterized by the temperature of $T_{\mathrm{dust}} \approx 700-1000$\,K and the blackbody radius of $R_{\mathrm{BB, dust}} \approx (2-3) \times 10^{16}$\,cm, corresponding to the luminosity of $L_{\mathrm{dust}} \approx 2 \times 10^{41}$\,erg\,s$^{-1}$. 

To estimate the dust mass, we also fit the IR excess with an optically-thin graybody model. In this case, the observed flux is
\begin{equation}
    f_{\lambda} = \frac{M_{\mathrm{dust}} B_{\lambda}(T_{\mathrm{dust}}) \kappa_{\lambda}}{D_{\mathrm{L}}^{2}},
    \label{eq:graybody}
\end{equation}
where $M_{\mathrm{dust}}$ is the dust mass and $\kappa_{\lambda}$ is the mass absorption coefficient of the dust grains. Adopting the opacity of 0.1-$\mu$m graphite grains, which is calculated using the optical constants in \citet{Zubko_1996_MNRAS_282_1321Z}, we obtain $M_{\mathrm{dust}} \approx (2-6) \times 10^{-3}~M_{\odot}$ with $T_{\mathrm{dust}} \approx 500$--$700$\,K at the three {\it NEOWISE} epochs; astronomical silicate provides a dust mass larger by a factor of $\sim 2$. Note that the graybody fit provides a lower dust temperature than the blackbody fit, because of the wavelength dependence of the absorption coefficient ($\kappa_{\lambda} \propto \lambda^{-1.4}$ for carbonaceous dust in the IR regime), and not a physically distinct component. The blackbody fit provides the minimum emitting radius (i.e., the radius in the optically thick limit), while the graybody fit is used to estimate the dust mass.

The dust component evolves non-monotonically over the {\it NEOWISE} epochs. At 166 days, it is relatively cool ($\sim 650$\,K) and extended ($\sim 3.0\times10^{16}$\,cm), while the blackbody fit at 331 days favors a hotter ($\sim 1000$\,K) and more compact ($\sim 1.6\times10^{16}$\,cm) component. For comparison, the freely expanding ejecta radius grows from $\sim 1\times10^{16}$\,cm at 166\,days (for $v_{\mathrm{ej}} \approx 8500$\,km\,s$^{-1}$), remaining smaller than $R_{\mathrm{BB,dust}}$ at the first epoch but becoming comparable to the dust radius at later phases. This non-monotonic temperature and radius evolution is difficult to explain with a single SN-heated dust component, but instead suggests contributions from both pre-existing dust and newly formed dust in the cold dense shell (CDS; Section~\ref{sec:discussion}).

\subsection{Late-time Optical Spectrum} \label{subsec:modeling_spec}

The broad, flat-topped H$\alpha$ profile with red-side suppression (Figure~\ref{fig:optical}c) is characteristic of emission from an expanding, detached shell that is partially attenuated by internal absorption. To quantify the emitting geometry, we fit the profile with a line-profile model for a geometrically thick, spherically symmetric shell, described in Appendix~\ref{app:line_profile}. The model shown in Figure~\ref{fig:optical}c adopts $V_{\max} = 8500$\,km\,s$^{-1}$, $R_{\mathrm{in}} = 4.5\times10^{16}$\,cm, $R_{\mathrm{out}} = 9.0\times10^{16}$\,cm (a thickness ratio of $R_{\mathrm{in}}/R_{\mathrm{out}} = 0.5$), a flat emissivity profile ($q = 0$), and a shell optical depth of $\tau_{\mathrm{shell}} = 0.3$. This model reproduces both the broad, flat-topped shape and the attenuation of the red wing of the H$\alpha$ emission.

These parameters indicate that the line-emitting region is distributed in a geometrically thick spherical shell subject to internal absorption, plausibly by dust associated with the interaction region, as inferred for other interacting SNe \citep[e.g.,][]{Smith_2008_ApJ_686_467S, Maeda_2013_ApJ_776_5M, Gall_2014_Nature_511_326G, Niculescu-Duvaz_2022_MNRAS_515_4302N}. The inferred shell radius of $\sim (4$--$9) \times 10^{16}$\,cm is consistent with the expected location of the ejecta and the CDS, $R_{\mathrm{ej}} \approx 9 \times 10^{16}$\,cm at 1271\,days for $v_{\mathrm{ej}} \approx 8500$\,km\,s$^{-1}$, and coincides with the dust-emitting radius derived from the SED at late phases (Section~\ref{subsec:modeling_sed}). This spatial coincidence supports a picture in which the H$\alpha$ emission, the CDS, and the dust all arise from the same ejecta--CSM interaction region.

\subsection{Bolometric Light Curves} \label{subsec:modeling_Lbol}

\begin{figure*}[htb]
\centering
\epsscale{1.17}

\gridline{
  \fig{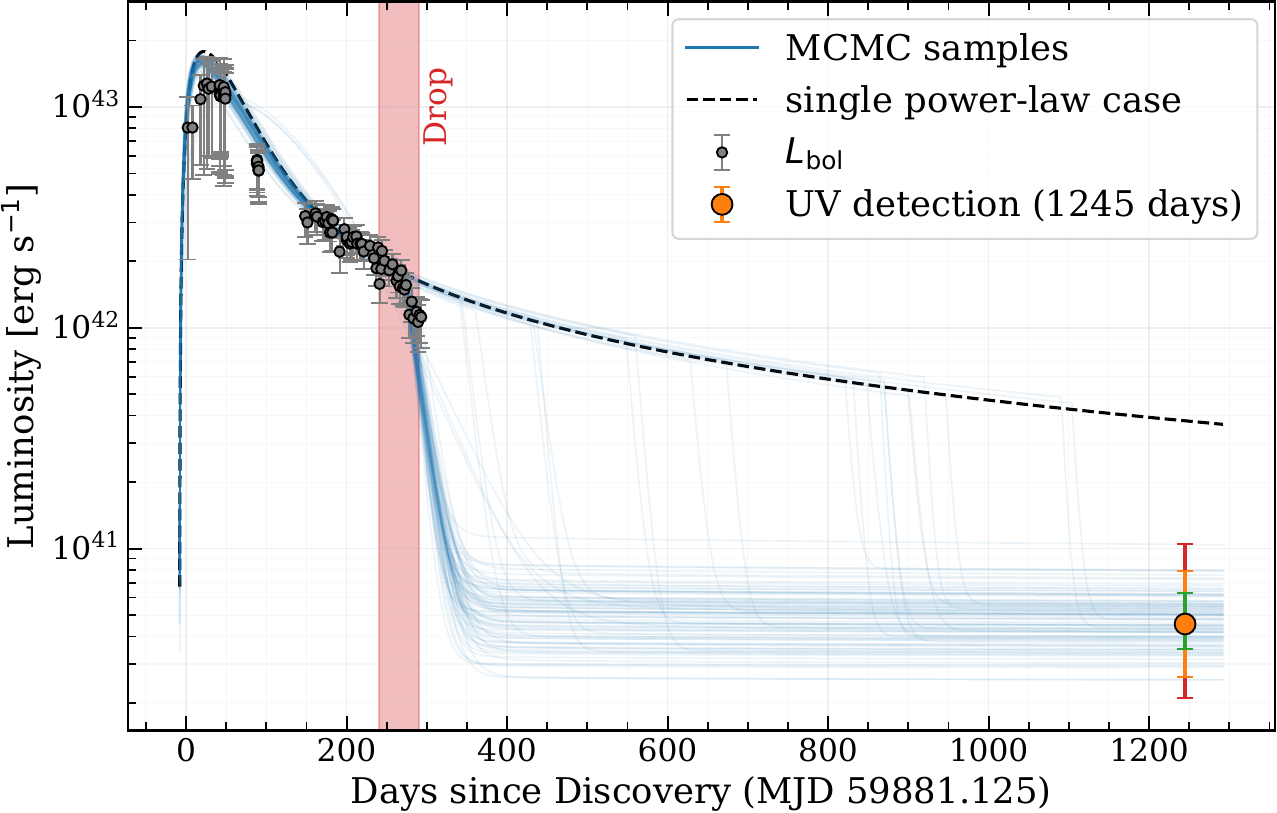}{0.49\textwidth}{}
  \fig{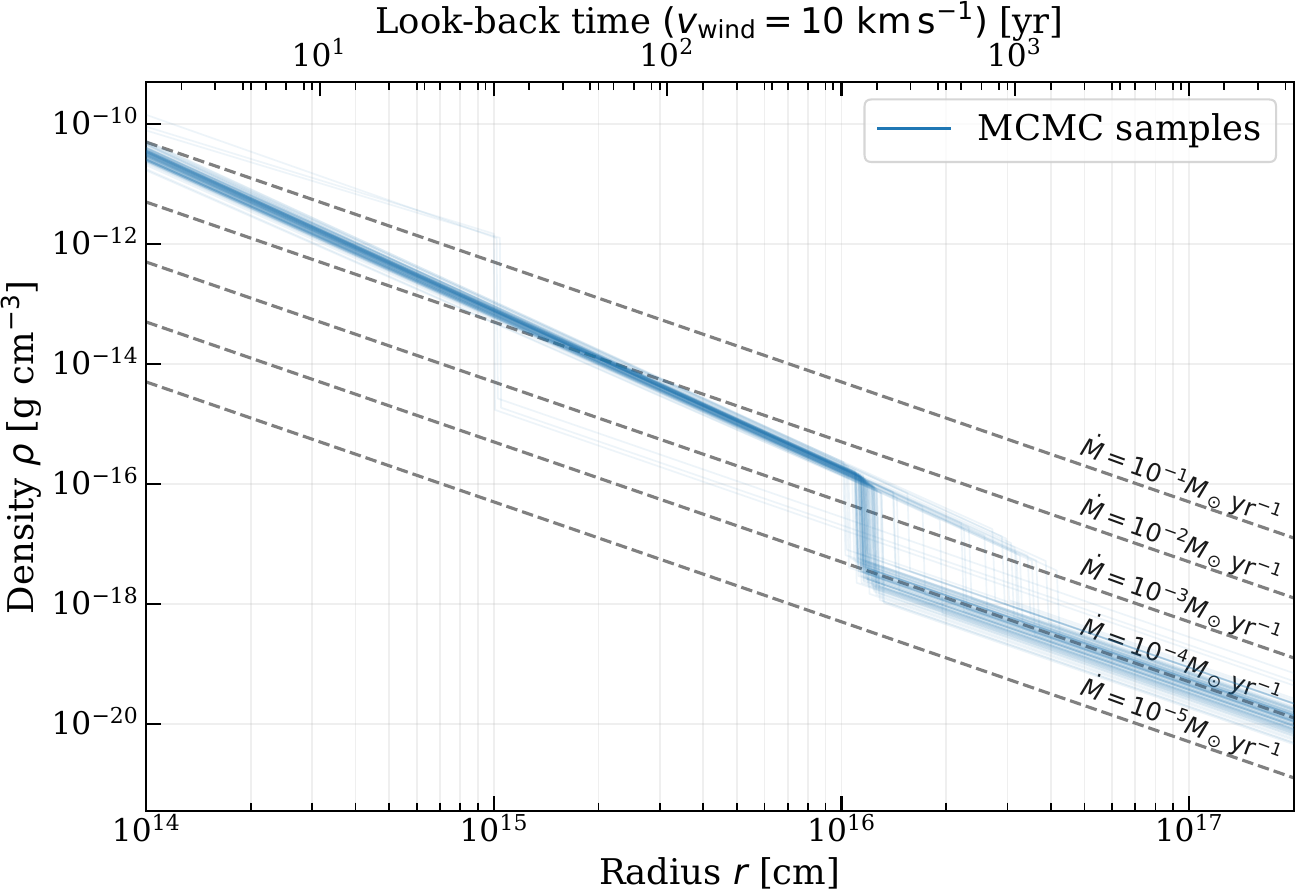}{0.5\textwidth}{}
}
\caption{
Left: Bolometric light curve of SN 2022yyz compared with the SN--CSM interaction model. The gray circles show the bolometric luminosities derived from blackbody fits to the $g$- and $r$-band photometry, adopting an intrinsic color of $(g-r)_{\rm int}=0.0$\,mag for the host-galaxy extinction correction. The error bars span the cases $(g-r)_{\rm int}=\pm0.1$\,mag. The orange circle is the bolometric luminosity at 1245 days after discovery, converted from the UV detection with a fraction of $f_{\rm UV}=0.8$. Its orange error bar shows the host-extinction uncertainty for the fiducial fraction, whereas the red and green bars show the most optimistic and the most pessimistic combinations of the host-extinction correction and of $f_{\rm UV}$ over the range $0.6$–$1.0$. The blue lines are 100 MCMC samples using the light curve model described in Appendix \ref{app:lc}, and the black dashed line shows the best fit model with a single power-law CSM distribution as a comparison. The red-shaded region shows the light curve drop around 300 days.
Right: CSM density profiles corresponding to the same 100 posterior samples shown in the left panel. The gray dashed lines are the profiles expected from steady mass loss ($s=2$), for the labeled mass-loss rates and wind velocity of $v_{\rm w}=10$\,km\,s$^{-1}$. The upper axis gives the corresponding look-back time $t=r/v_{\rm w}$ before the explosion for the wind velocity.
}
\label{fig:lc_model}
\end{figure*}

Figure~\ref{fig:lc_model} shows the bolometric light curve of SN\,2022yyz, which is derived from blackbody fits to the $g$- and $r$-band photometry. Note that the $g-r$ color used for the optical blackbody estimate may be affected by H$\alpha$ emission in the $r$ band, particularly at late phases. The peak luminosity of $1.3^{+0.7}_{-0.4}\times10^{43}$\,erg\,s$^{-1}$ (Section~\ref{subsec:obsprop_optical}) is difficult to reconcile with radioactive decay: a simple one-zone estimate \citep{Arnett_1982_ApJ_253_785A} requires a $^{56}$Ni mass of $M_{\mathrm{Ni}} \gtrsim 1~M_{\odot}$, which is an order of magnitude larger than that of canonical SNe II \citep[$\lesssim 0.1~M_{\odot}$; e.g.,][]{Anderson_2019_AA_628_A7A}. Furthermore, the light curve does not exhibit a clear plateau, suggesting that hydrogen recombination does not significantly contribute to powering the light curve. Such a high luminosity instead requires an additional power source, most naturally CSM interaction; combined with the direct UV signature at late times (Section~\ref{subsec:obsprop_uv}), this leads us to reinterpret SN\,2022yyz as an interaction-powered event despite its early Type II classification.

The bolometric light curve exhibits two distinct phases: a luminous, monotonically declining early peak, and a late-time UV luminosity at 1245 days that lies about two orders of magnitude fainter than peak, with a drop at $\sim 300$\,days. In the interaction picture, this drop marks the epoch at which the forward shock sweeps through a dense inner CSM component; the subsequent, much fainter emission traces continued interaction with a lower-density outer CSM. This behavior cannot be reproduced by a single power-law CSM density profile (black dashed line in Figure~\ref{fig:lc_model}), which fails to match the early peak and the late-time UV point simultaneously. We therefore model the CSM with a broken power-law density profile, with a density jump at a drop radius $r_{\mathrm{drop}}$; the full formalism is described in Appendix~\ref{app:lc}.

We fit the bolometric light curve using a Markov Chain Monte Carlo (MCMC) method with six free parameters: the inner and outer CSM masses $M_{\mathrm{CSM,in}}$ and $M_{\mathrm{CSM,out}}$, the inner density slope $s_{1}$, the drop radius $r_{\mathrm{drop}}$, the ejecta mass $M_{\mathrm{ej}}$, and the explosion energy  $E_{\mathrm{ej}}$. The outer slope is fixed to $s_{2} = 2$, appropriate for steady mass loss. The inner and outer radii are fixed to $r_{\mathrm{in}} = 1\times 10^{13}$\,cm and $r_{\mathrm{out}} = 2\times10^{17}$\,cm, which are well inside and outside the region probed by the light curve and does not affect the results. For the conversion efficiency from shock kinetic energy to radiation, we adopt $\varepsilon = 0.2$, following numerical calculations of interacting SNe, which find $\varepsilon \approx 0.1$--$0.3$
\citep{Moriya_2013_MNRAS_428_1020M}.

The posterior parameters from the MCMC fit are summarized in Table~\ref{tab:mcmc}. The inferred ejecta mass is consistent with the explosion of a slightly stripped massive star. The model reproduces both the luminous, linearly declining early light curve and the late-time luminosity derived from the UV detection: the early emission is powered by interaction with the dense inner CSM ($\rho \propto r^{-2.6}$ within $r_{\mathrm{drop}}$), while continued interaction of the forward shock with the extended outer, wind-like CSM sustains the late-time emission and produces the plateau-like evolution following the drop.

\begin{deluxetable}{lccc}
\tabletypesize{\footnotesize}
\tablewidth{0pt}
\tablecaption{Priors and posterior parameters of the SN--CSM interaction model.
\label{tab:mcmc}}
\tablehead{
  \colhead{Parameter} & \colhead{Prior\tablenotemark{a}} & \colhead{Posterior\tablenotemark{b}} & \colhead{Unit}
}
\startdata
$M_{\rm CSM,in}$   & [0.1,10]  & $2.72^{+0.65}_{-0.35}$   & $M_\odot$       \\
$M_{\rm CSM,out}$  & [0.1,10]  & $0.50^{+0.21}_{-0.14}$   & $M_\odot$       \\
$s_{1}$            & [1.1,2.9] & $2.62^{+0.03}_{-0.03}$   &                 \\
$r_{\rm drop}$     & [0.1,10]     & $1.21^{+0.65}_{-0.09}$   & $10^{16}$\,cm   \\
$M_{\rm ej}$       & [10,20]   & $10.77^{+0.92}_{-0.56}$  & $M_\odot$       \\
$E_{\rm ej}$       & [0.5,30]  & $2.63^{+0.24}_{-0.24}$   & $10^{51}$\,erg  \\
\enddata
\tablenotetext{a}{The prior ranges are shown in physical units. Uniform priors are adopted in logarithmic space for $M_{\rm CSM,in}$, $M_{\rm CSM,out}$, $r_{\rm drop}$, and $E_{\rm ej}$, and in linear space for $s_1$ and $M_{\rm ej}$.}
\tablenotetext{b}{Median and $16$th--$84$th percentiles of the marginalized posterior distributions. The parameter uncertainties are statistical uncertainties within the adopted model.}
\end{deluxetable}

The derived CSM structure can be translated into a mass-loss history of the progenitor. Assuming a constant wind velocity of $v_{\mathrm{w}} = 10$\,km\,s$^{-1}$ (typical for red supergiants; RSGs), the drop radius corresponds to a lookback time of a few hundred years. The effective mass-loss rate, $\dot{M}(r) = 4\pi r^{2} \rho_{\mathrm{CSM}}(r)\,v_{\mathrm{w}}$, is $\sim 10^{-2}~M_{\odot}$\,yr$^{-1}$ within the inner CSM and increases toward smaller radii (i.e., toward the explosion epoch), and $\sim 10^{-4}~M_{\odot}$\,yr$^{-1}$ in the outer CSM. This requires that the progenitor of SN\,2022yyz underwent an enhancement of its mass-loss rate by roughly two orders of magnitude during the final few hundred years before core collapse.


\section{Discussion and Conclusions} \label{sec:discussion}

\begin{figure*}[htb]
\centering
\epsscale{1.0}
\plotone{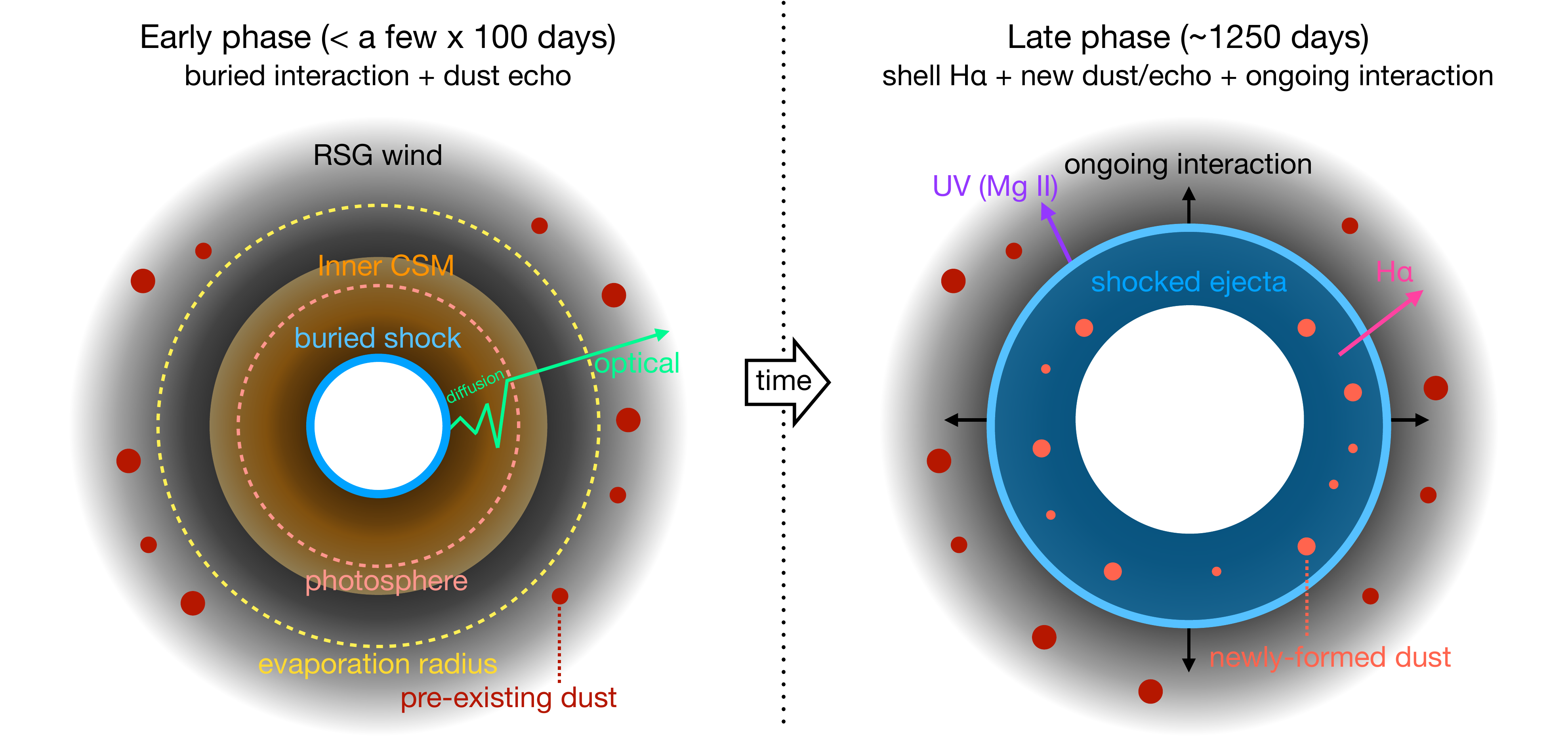}
\caption{
Schematic picture of SN 2022yyz at early ($\lesssim$ a few $\times 10^{2}$\,days; left) and late ($\sim$1250\,days; right) phases. 
Left: The forward shock (blue) is initially embedded in the dense inner CSM (orange shading), which is optically thick to electron scattering. The interaction luminosity therefore escapes as thermalized, diffused optical radiation (green arrow), producing the luminous light curve. Part of this radiation is absorbed and re-emitted as the MIR echo by pre-existing dust grains (red dots) in the outer CSM beyond the dust evaporation radius (yellow).
Right: The shock has swept up the inner CSM and formed a geometrically-thick shocked shell (blue shading), which produces the flat-topped hydrogen emission profile. The newly-formed dust within the CDS (coral dots) attenuates the red side of the line and contributes to the late-time IR emission, together with the echo through the pre-existing dust heated by the ongoing interaction. The continued interaction of the forward shock with the outer wind powers the UV (\ion{Mg}{2}) emission.
}
\label{fig:schematic_picture}
\end{figure*}

We have presented a multi-wavelength study of SN\,2022yyz, a nearby CCSN that was selected through its late-time MIR excess in {\it NEOWISE}. Although SN\,2022yyz was first classified as a Type II SN, our follow-up observations lead us to reclassify it as a luminous interacting SN -- that remains powered by CSM interaction $\sim 3.5$ years after discovery. The source is detected in late-time {\it HST}/WFC3 imaging in F275W, implying UV line emission with $L \approx 3.6\times10^{40}$\,erg\,s$^{-1}$, corroborating a drop in the bolometric light curve inferred from optical data at $\sim 300$\,days. A Keck/LRIS spectrum at 1271 days shows a broad, flat-topped H$\alpha$ profile with an attenuated red wing, as expected for a quasi-spherical shocked shell containing dust. A semi-analytic interaction model that simultaneously reproduces the early optical light curve and this late-time UV luminosity requires $\sim 2.7~M_{\odot}$ of inner, dense CSM within $\sim 10^{16}$\,cm surrounded by wind-like outer CSM with $\dot{M} \sim 10^{-4}\,{\rm M_\odot\,yr^{-1}}$ ($\sim 0.5\,M_\odot$ within $r_{\mathrm{out}}$). Although these estimates are model dependent, the total inferred CSM mass of SN~2022yyz  exceeds that inferred for Type IIP SNe; e.g., SN~2023ixf \citep[$\lesssim 0.1\,M_{\odot}$;][]{JacobsonGalan_2023_ApJL_954_L42J}, and is more comparable to that of long-lived Type II; e.g., SN~2021irp \citep[a few $M_{\odot}$;][see below]{Reynolds_2025_AA_702_A213R}, while remaining below the value inferred from Type IIn SNe; e.g., SN~2010jl \citep[$\sim 10\,M_{\odot}$;][]{Ofek_2014_ApJ_781_42O}. Thus, SN\,2022yyz demonstrates our strategy: MIR excess selects events with ongoing CSM interaction, and UV photometry combined with late-time spectroscopy constrains their quantitative mass-loss history. In the following, we discuss what this reconstructed history implies for the nature of the interaction, the mass-loss mechanism, and the associated dust.

\subsection{The Nature of the CSM Interaction}

SN\,2022yyz reaches a bright peak luminosity of $\sim 1\times10^{43}$\,erg\,s$^{-1}$ and then declines monotonically. As shown in Section~\ref{subsec:modeling_Lbol}, radioactive decay cannot power this luminosity, leaving CSM interaction as the most natural energy source. Reproducing the full data set, however, requires a broken power-law CSM rather than a single power law, for three complementary observational signatures: (i) the early optical spectrum is nearly featureless, indicating a buried shock within an optically thick inner CSM; (ii) the bolometric light curve drops after $\sim 300$\,days and then connects to the faint late-time UV detection, implying a decline in the interaction power as the shock crosses a density break; and (iii) the {\it NEOWISE} colors evolve blueward across this transition, consistent with a hot CDS dust component becoming exposed.

The fitted inner CSM ($M_{\mathrm{CSM,in}} \approx 2.7~M_{\odot}$, $\rho \propto r^{-2.6}$) implies that the electron-scattering photosphere ($\tau = 1$) lies at $\sim 5\times10^{15}$\,cm, so that even after the breakout, the forward shock remains buried below the photosphere (Figure~\ref{fig:schematic_picture}). In this regime, the light curve is governed by photon diffusion and shock dynamics \citep{Chevalier_2011_ApJL_729_6C}: the $\sim 30$-day rise matches the diffusion timescale, while the quasi-linear decline reflects the deceleration of the shock. The luminous, IIL-like appearance of SN\,2022yyz is therefore a natural consequence of interaction buried in a compact, dense CSM \citep[e.g.,][]{Khatami_2024_ApJ_972_140K}.

\subsection{The Origin of the Terminal Mass Loss}

The compact and massive CSM requires an intense mass-loss rate of $\gtrsim 10^{-2}~M_{\odot}$\,yr$^{-1}$ $(v_{\mathrm{w}}/10 {\mathrm{~km~s^{-1}}})$ sustained over the final few hundred years. The gravitational energy required to unbind such a mass from the surface of an RSG is
\begin{equation}
    E_{\mathrm{bind}}\sim 3\times 10^{47}~\mathrm{erg}
    \left(\frac{M_{\mathrm{RSG}}}{15M_{\odot}}\right)
    \left(\frac{M_{\mathrm{CSM}}}{3M_{\odot}}\right)
    \left(\frac{R_{\mathrm{RSG}}}{500R_{\odot}}\right)^{-1}.
\end{equation}
While this energy is comparable to or smaller than the wave energy that may be deposited in the envelope during the final nuclear-burning stages \citep{Quataert_2012_MNRAS_423L_92Q}, the wave-driven model faces two difficulties. First, even under optimistic efficiencies, standard RSG models eject $\lesssim 1~M_{\odot}$ by this mechanism \citep{Fuller_2017_MNRAS_470_1642F}, short of the required mass of $\sim 2.7~M_{\odot}$. Second, and more fundamentally, the wave-driven model is tied to the late nuclear-burning timescales of $\lesssim 10$\,yr, whereas the inferred mass loss must begin $\gtrsim 100$\,yr before explosion. Pulsation-driven superwinds can, in principle, approach the required rate \citep{Yoon_2010_ApJL_717_62Y}, but only near their theoretical upper limit; adopting a faster wind would increase the inferred mass-loss rate and make this explanation more challenging. Thus, such single-star mechanisms are not excluded, but would require an unusually extreme and sustained eruption.

A more natural alternative scenario is partial envelope ejection through binary interaction, i.e., Roche-lobe overflow or common-envelope (CE) ejection \citep[e.g.,][]{Chevalier_2012_ApJL_752_2C, Metzger_2017_MNRAS_471_3200M}. CE ejecta move at roughly the system escape velocity ($\sim 50$--$100$\,km\,s$^{-1}$ for an RSG envelope), corresponding to lookback times of decades to centuries. Importantly, this scenario bypasses the timing problem of the single-star models: binary evolution calculations generically place the onset of the CSM density enhancement at $\sim 10^{16}$\,cm, set by the binary separation rather than by the nuclear-burning timescale \citep{Ercolino_2024_AA_685_A58E, Ercolino_2026_AA_706_A169E, Matsuoka_2024_ApJ_963_105M, Tsai_2026_ApJ_1005_L36T}, comparable to the drop radius we infer. This suggests that the drop may be a signature of a binary origin, and a more violent CE ejection could further account for the large inner CSM mass that stable mass transfer alone cannot supply. The timing itself is also not necessarily fine-tuned: wave-driven energy deposition during late core burning can inflate the envelope and trigger Roche-lobe overflow or CE evolution shortly before core collapse \citep{Mcley_2014_MNRAS_445_2492M}.

A potential concern for the binary scenario is the CSM geometry. Mass loss through binary evolution is expected to concentrate toward the orbital plane \citep[e.g.,][]{Scherbak_2025_ApJ_990_172S}, and SN\,2021irp, which is an event resembling SN\,2022yyz, shows high continuum polarization indicative of a disk-like CSM \citep{Reynolds_2025_AA_702_A213R, Nagao_2025_AA_699_A283N}. In contrast, the flat-topped H$\alpha$ profile of SN\,2022yyz favors a quasi-spherical emitting region and disfavors a geometrically thin disk. However, this does not exclude a CE origin: long-term simulations show that the asymmetric ejecta produced during dynamical inspiral radially expand and evolve toward shell-like geometry within decades, while any equatorial circumbinary structure from prior Roche-lobe overflow can be substantially disrupted by the CE event \citep{Reichardt_2019_MNRAS_484_631R, Iaconi_2019_MNRAS_489_3334I}. Although these calculations mostly consider low-mass giant binaries rather than
massive RSG progenitors, they demonstrate that a quasi-spherical, large-scale interaction region is not incompatible with a CE origin.

\subsection{Dust Properties and the Infrared Emission}

The inferred CSM structure is broadly consistent with the IR emission. The outer CSM of $\sim 0.5~M_{\odot}$ and the graybody dust mass of $(2$--$6)\times10^{-3}~M_{\odot}$ imply a dust-to-gas ratio of $(4$--$12)\times10^{-3}$. As shown in Section~\ref{subsec:modeling_sed}, the early MIR emission arises from pre-existing dust lying outside the ejecta, and its blackbody radius at 166 days ($\sim 3\times10^{16}$\,cm) is comparable to the dust evaporation radius of a few $\times10^{16}$\,cm set by the peak luminosity $L_{\mathrm{peak}} \approx 1\times10^{43}$\,erg\,s$^{-1}$ in thermal equilibrium. The early MIR excess is thus naturally explained as an echo by the pre-existing dust outside the sublimation radius.

The blueward color evolution of the dust component described in Section~\ref{subsec:modeling_sed} may indicate two MIR emission components. The initial large, cool component is the pre-existing CSM dust heated by the SN, whereas the subsequent hot, compact phase may reflect an increasing contribution from newly formed dust in the CDS behind the shock. This interpretation is reinforced by energetics: the late-time IR emission cannot be sustained by an echo alone. The characteristic echo duration is bounded by the light-crossing time of the CSM: $2r_{\mathrm{out}}/c \lesssim 150$\,days, whereas the MIR luminosity remains nearly constant beyond $\gtrsim 500$\,days. Therefore, the late-time IR requires continued heating of surviving pre-existing dust by the ongoing interaction ($L \sim 10^{41}$\,erg\,s$^{-1}$) and/or emission from newly condensed hot dust in the CDS. Furthermore, the red-side attenuation of the broad H$\alpha$ profile independently supports the presence of newly formed dust associated with the CDS (Section~\ref{subsec:modeling_spec}).

Energetically, the reprocessed MIR energy accumulated so far, $E_{\mathrm{rad,MIR}} \approx 2\times10^{49}$\,erg, accounts for $\sim 10\%$ of the optical radiated energy $E_{\mathrm{rad,opt}} \approx 2\times10^{50}$\,erg -- a lower limit, because this fraction may grow as the interaction continues. Since the pre-existing dusty CSM is optically thick to optical radiation, the light absorbed by the dust is reprocessed into the IR, and the reprocessed fraction measures the dust covering fraction, $f_{\mathrm{cov}} \gtrsim 0.1$, implying clumpy dust as seen in RSG winds \citep{Humphreys_2007_AJ_133_2716H, Scicluna_2015_AA_584_L10S}.

\subsection{Summary}

Although IR searches for CSM interaction have been historically limited by synoptic search capabilities, SN\,2022yyz demonstrates the end-to-end methodology of our ongoing {\it NEOWISE}-selected follow-up campaign. In particular, the late-time UV line emission inferred from our data corroborates strong CSM interaction as the origin of luminous late-time IR emission previously suggested from late-time optical spectroscopy \citep{Myers_2024_ApJ_976_230M}. Given the extremely faint late-time optical and UV emission which has thus far prevented a precise identification for this object, the luminous IR emission (that is opaque to the high energy CSM interaction radiation) has therefore proven to be a powerful beacon to probe late-stage mass loss in the massive star progenitor. 

Our analysis reveals a progenitor that ejected $\sim 2.7~M_{\odot}$ in the last few hundred years before core collapse -- an eruption pointing to a possible common-envelope ejection in a binary system. The forthcoming full-sample analysis, combining the approved {\it HST} UV imaging and Keck/LRIS spectroscopy of MIR-selected CCSNe across multiple spectroscopic subtypes, will quantify the prevalence, mass and energy budgets, and geometries of such envelope-scale terminal mass loss, enabling systematic comparison against binary-interaction and single-star (e.g., wave-driven and pulsational-instability) predictions. Our program serves as a complement to the high sensitivity surveys of the {\it Rubin} observatory, that may soon provide routine detection of the very faint late-time optical/UV emission \citep{Dessart2022, Terwel2025}. In the IR landscape, our program will set the stage for the synergistic utilization of the {\it Roman} Space Telescope surveys \citep{Rose_2021arXiv211103081R} with planned UV initiatives such as ULTRASAT \citep{Shvartzvald_2024_ApJ_964_74S} and UVEX \citep{Kulkarni_2021arXiv211115608K} later in the decade.

\begin{acknowledgments}

Based on observations made with the NASA/ESA Hubble Space Telescope, obtained at the Space Telescope Science Institute, which is operated by the Association of Universities for Research in Astronomy, Inc., under NASA contract NAS5-26555. These observations are associated with program GO-18016. The data presented herein were also obtained at Keck Observatory (2026A/N194 and 2026B/N150), which is a private 501(c)3 non-profit organization operated as a scientific partnership among the California Institute of Technology, the University of California, and the National Aeronautics and Space Administration. The Observatory was made possible by the generous financial support of the W. M. Keck Foundation. The authors wish to recognize and acknowledge the very significant cultural role and reverence that the summit of Maunakea has always had within the Native Hawaiian community. We are most fortunate to have the opportunity to conduct observations from this mountain. This paper includes data gathered with the 6.5 meter Magellan Telescopes located at Las Campanas Observatory, Chile.

K.U. acknowledges financial support from the Japan Society for the Promotion of Science Overseas Fellowship, and from Keck PI Data Award managed by NExScI for NASA. We acknowledge the support of the National Aeronautics and Space Administration through ADAP grant number 80NSSC24K0663. V.K. was supported by NASA through the NASA Hubble Fellowship grant \#HST-HF2-51578.001-A awarded by the Space Telescope Science Institute, which is operated by the Association of Universities for Research in Astronomy, Inc., for NASA, under contract NAS5-26555. G.M. is supported by the Brinson Foundation through the Brinson Prize Fellowship Program.

\end{acknowledgments}

\facilities{WISE(NEOWISE), HST(WFC3), Keck:I(LRIS), Magellan:Baade(FourStar)}

\software{
          astropy \citep{astropy:2013, astropy:2018, astropy:2022},
          PypeIt \citep{pypeit:joss_pub},
          NumPy \citep{Numpy},
          Matplotlib \citep{Matplotlib},
          SciPy \citep{Scipy},
          emcee \citep{emcee_1}
          }

\bibliography{manuscript}{}
\bibliographystyle{aasjournal}


\appendix

\section{Photometry tables} \label{app:photo_tables}

The photometry of SN\,2022yyz is summarized in Table~\ref{tab:phot}.

\begin{deluxetable}{ccllc}
\tablewidth{0pt}
\tablecaption{Photometry of SN\,2022yyz\label{tab:phot}}
\tablehead{
  \colhead{MJD} &
  \colhead{Phase\tablenotemark{a}} &
  \colhead{Telescope/Instrument} &
  \colhead{Filter} &
  \colhead{mag(error) \tablenotemark{b}} 
}
\startdata
59850.16 & $-31.0$  & WISE/NEOWISE           & $W1$    & $>19.28$ \\
59850.16 & $-31.0$  & WISE/NEOWISE           & $W2$    & $>18.91$ \\
60047.46 & $166.3$  & WISE/NEOWISE           & $W1$    & $16.82 (0.02)$ \\
60047.46 & $166.3$  & WISE/NEOWISE           & $W2$    & $16.55 (0.02)$ \\
60211.68 & $330.6$  & WISE/NEOWISE           & $W1$    & $16.60 (0.02)$ \\
60211.68 & $330.6$  & WISE/NEOWISE           & $W2$    & $16.43 (0.02)$ \\
60413.06 & $531.9$  & WISE/NEOWISE           & $W1$    & $16.74 (0.02)$ \\
60412.99 & $531.9$  & WISE/NEOWISE           & $W2$    & $16.39 (0.02)$ \\
60808.3  & $927.2$  & Magellan/FourStar      & $J$     & $20.41 (0.03)$ \\
60808.3  & $927.2$  & Magellan/FourStar      & $H$     & $19.96 (0.12)$ \\
60808.3  & $927.2$  & Magellan/FourStar      & $K_{s}$ & $18.61 (0.11)$ \\
61126.0  & $1244.9$ & \textit{HST}/WFC3-UVIS & F225W   & $>26.1$ \\
61126.0  & $1244.9$ & \textit{HST}/WFC3-UVIS & F275W   & $25.60 (0.15)$ \\
61193.4  & $1312.3$ & Magellan/FourStar      & $H$     & $20.27 (0.11)$ \\
61193.4  & $1312.3$ & Magellan/FourStar      & $K_{s}$ & $19.35 (0.06)$ \\
\enddata
\tablenotetext{a}{Days since discovery (MJD 59881.125)}
\tablenotetext{b}{Observed magnitudes in the AB system uncorrected for Milky Way or host-galaxy extinction. Limits are $5\sigma$.}
\end{deluxetable}

\onecolumngrid
\section{Model Formulation} \label{app:model}

\subsection{H$\alpha$ Line-Profile Modeling} \label{app:line_profile}

To interpret the H$\alpha$ profile of SN 2022yyz, we model the emission from an expanding, geometrically thick shell following the line-profile formalism of \citet{Jerkstrand_2017_snhandbook_795J}. We consider a spherically symmetric shell extending from an inner radius $R_{\mathrm{in}}$ to an outer radius $R_{\mathrm{out}}$, in homologous expansion. At an epoch $t$, the velocity field is $V(r) = r/t = V_{\max}(r/R_{\mathrm{out}})$, with $V_{\max} = R_{\mathrm{out}}/t$ and $V_{\min} = R_{\mathrm{in}}/t$.

We adopt cylindrical velocity coordinates $(v_{p}, v_{\mathrm{los}})$, where $v_{\mathrm{los}} = z/t$ is the line-of-sight velocity (with the $z$-axis directed toward the observer) and $v_{p} = p/t$ is the component in the plane of the sky. We assume a radial, line-integrated volume emissivity $j(V) = j_{0}\,(V/V_{\max})^{q}$ within the shell. At each observed velocity, only the isovelocity plane $v_{\mathrm{los}} = -v_{\mathrm{obs}}$ contributes, where positive $v_{\mathrm{obs}}$ denotes a receding (redshifted) velocity. Neglecting scattering, the emergent profile is
\begin{equation}
    F(v_{\mathrm{obs}}) \propto
    \int_{\max\!\left(V_{\min},\,|v_{\mathrm{obs}}|\right)}^{V_{\max}}
    j(V)\,e^{-\tau(V,v_{\mathrm{obs}})}\,V\,dV,
    \label{eq:profile}
\end{equation}
where the lower integration limit selects either the inner edge of the shell or the isovelocity plane itself when $|v_{\mathrm{obs}}|$ exceeds $V_{\min}$, and $\tau(V, v_{\mathrm{obs}})$ is the optical depth from the emitting point to the observer along the line of sight.

We assume a uniform absorption coefficient within the shell, $\alpha_{\mathrm{abs}} = \tau_{\mathrm{shell}} / (R_{\mathrm{out}} - R_{\mathrm{in}})$, and set it to zero in the inner cavity. The line-of-sight optical depth $\tau(V, v_{\mathrm{obs}})$ is obtained by integrating $\alpha_{\mathrm{abs}}$ over the shell material between the emitting point and the outer boundary along the line of sight. Emission from the far (red) side of the shell traverses a larger absorbing column than that from the near (blue) side, so the far-side emission is preferentially attenuated, producing the observed suppression of the red wing. In the optically thin limit ($\tau_{\mathrm{shell}} \rightarrow 0$), the profile reduces to the standard flat-topped profile of a uniformly emitting spherical shell.

\subsection{SN--CSM Interaction Light-Curve Modeling} \label{app:lc}

To estimate the CSM structure and interaction energetics, we model the bolometric light curve with the semi-analytic SN--CSM interaction model of \citet{Nagao_2020_MNRAS_497_5395N} and \citet{Uno_2023_ApJ_944_204U}. Here, we summarize the essential formalism; for full details, see \citet{Uno_2023_ApJ_944_204U}. The only modification that we introduce is the CSM density profile: as discussed in Section~\ref{subsec:modeling_Lbol}, a single power-law profile cannot reproduce both the early peak and the late-time UV luminosity, so we adopt the broken power-law of Equation~\ref{eq:csm}.

The CSM density is described by a double power-law confined to $[r_{\mathrm{in}}, r_{\mathrm{out}}]$:
\begin{equation}
    \rho_{\rm CSM}(r) =
    \begin{cases}
    D_{1} r^{-s_{1}}, & r_{\rm in} \le r < r_{\rm drop},\\
    D_{2} r^{-s_{2}}, & r_{\rm drop} \le r \le r_{\rm out},
    \end{cases}
    \label{eq:csm}
\end{equation}
where the normalizations $D_{1}$ and $D_{2}$ are set by the CSM masses $M_{\mathrm{CSM,in}}$ and $M_{\mathrm{CSM,out}}$ enclosed in the inner and outer regions; for example,
\begin{equation}
    D_{1} = \frac{(3-s_{1})\,M_{\mathrm{CSM,in}}}{4\pi\left(r_{\rm drop}^{3-s_{1}} - r_{\rm in}^{3-s_{1}}\right)}.
\end{equation}
The outer slope is fixed to $s_{2} = 2$, assuming the distant CSM is formed by steady mass loss.

The freely expanding ejecta follow the standard broken power-law profile \citep[e.g.,][]{Moriya_2013_MNRAS_435_1520M}:
\begin{equation}
    \rho_{\rm ej}(r,t)\propto
\begin{cases}
    t^{-3}(r/t)^{-n}, & v_{\mathrm{ej}} \ge v_t,\\
    t^{-3}(r/t)^{-\delta}, & v_{\mathrm{ej}} < v_t,
\end{cases}
    \quad v_{\mathrm{ej}} \equiv \frac{r}{t},
\end{equation}
with transition velocity
\begin{equation}
    v_t=\left[\frac{2(5-\delta)(n-5)E_{\rm ej}}
    {(3-\delta)(n-3)M_{\rm ej}}\right]^{1/2},
\end{equation}
where $E_{\mathrm{ej}}$ and $M_{\mathrm{ej}}$ are the ejecta kinetic energy and mass. We adopt $n=12$ and $\delta = 1$, appropriate for the explosion of an RSG \citep[][]{Matzner_1999_ApJ_510_379M}.

Assuming a geometrically thin shocked shell, we evolve the shell from the equation of motion:
\begin{align}
    M_{\mathrm{sh}}(t)\,\frac{dv_{\mathrm{sh}}(t)}{dt} &= 4\pi r_{\mathrm{sh}}^{2}(t)
    \Bigl[ \rho_{\mathrm{ej}}(r_{\mathrm{sh}},t)\left(v_{\mathrm{ej}}(r_{\mathrm{sh}},t)-v_{\mathrm{sh}}(t)\right)^{2} \nonumber\\
    &\quad - \rho_{\mathrm{CSM}}(r_{\mathrm{sh}})\left(v_{\mathrm{sh}}(t)-v_{\mathrm{CSM}}\right)^{2} \Bigr],
\end{align}
where $v_{\mathrm{CSM}}$ is the CSM velocity and the shell mass is the sum of
the swept-up ejecta and CSM:
\begin{equation}
    M_{\mathrm{sh}}(t) = \int_{r_{\mathrm{in}}}^{r_{\mathrm{sh}}(t)} 4\pi r^{2} \rho_{\mathrm{CSM}}(r)\,dr
    + \int_{r_{\mathrm{sh}}(t)}^{r_{\mathrm{ej,max}}(t)} 4\pi r^{2} \rho_{\mathrm{ej}}(r,t)\,dr.
\end{equation}

The luminosity dissipated at the forward shock is a fraction $\varepsilon$ of the incoming kinetic-energy flux,
\begin{equation}
    L_{\mathrm{sh}}(t) = \varepsilon\,\frac{dE_{\mathrm{kin}}(t)}{dt},
\end{equation}
with
\begin{equation}
    dE_{\mathrm{kin}}(t) = 4\pi r_{\mathrm{sh}}^{2}(t)
    \left( \tfrac{1}{2}\rho_{\mathrm{CSM}}(r_{\mathrm{sh}}(t))\,v_{\mathrm{sh}}^{2}(t) \right) dr.
\end{equation}
The conversion efficiency $\varepsilon$ depends on the ejecta-to-CSM mass ratio; we adopt $\varepsilon = 0.2$ as justified in Section~\ref{subsec:modeling_Lbol}.

Finally, the observed light curve accounts for photon diffusion. The effective optical depth is
\begin{equation}
    \tau_{\mathrm{diff}} = \frac{\kappa_{\mathrm{es}} M_{\mathrm{sh}}}{4\pi r_{\mathrm{sh}}^2}
    + \int_{r_{\mathrm{sh}}(t)}^{r_{\mathrm{out}}} \kappa_{\mathrm{es}} \rho_{\mathrm{CSM}}\,dr,
\end{equation}
where $\kappa_{\mathrm{es}} = 0.34$ cm$^{2}$ g$^{-1}$ is the electron-scattering opacity of the ionized gas. The emergent luminosity is
\begin{equation}
    L(t) = \int_{0}^{t} \frac{L_{\mathrm{sh}}(t^{\prime})}{t_{\mathrm{diff}}(t^{\prime})}
    \exp\!\left( -\frac{t-t^{\prime}}{t_{\mathrm{diff}}(t^{\prime})} \right) dt^{\prime},
\end{equation}
with
\begin{equation}
    t_{\mathrm{diff}}(t) = \frac{\tau_{\mathrm{diff}}(t)\,r_{\mathrm{sh}}(t)}{c}.
\end{equation}

\end{document}